**Cleavage toughness of single crystals**

Faming Gao

School of Chemical Engineering and Materials Science, Tianjin University of Science and Technology, Tianjin 300457, China

Griffith and Irwin theories encounter difficulty in analyzing cleavage phenomenon from atomic level. Here, a new definition of the fracture toughness is proposed from the point of view of chemical bond. We find that the fracture toughness can be defined by the bond strength (the appropriate elastic modulus ) and the bond density. This appropriate elastic modulus of single crystals is obtained using the complex variable function method. The calculated results of cleavage toughness and surface energy of typical ionic and covalent crystals by the present formulae are in excellent agreement with the experimental values. It demonstrates that our method offers a concise tool for predicting the cleavage toughness, the energy release rate and the surface energy of crystal cleavage planes.

Under the action of a force the breaking of a solid into pieces is a common phenomenon in engineering and in nature [1, 2]. One of most fascinating fracture phenomenon is the cleavage of single crystals. Cleavage also play an important role in the gemstone industry, the crushing of minerals and the industrial abrasives field. The cleavage fracture is a form of brittle fracture. Specifically, the cleavage fracture surface is exhibiting a mirror-like appearance. Researchers are dedicated to uncovering what determines the crystal plane along which the cleavage occurs. The ionic charge, surface free energy, bond density and elastic modulus have all been attempted as criterions for cleavage [3]. But each cannot, by itself, consistently determine the cleavage planes. Schultz et al. experimentally investigated single crystal cleavage, and suggested the measured cleavage toughness as the criterion of single crystal cleavage [3]. However, the feasible corresponding theoretical expression for single crystal cleavage toughness is still being explored.

In the macroscopic aspect, Griffith proposed a fracture criterion based on a thermodynamic energy balance [4]. In 1921, Griffith proposed that when a crack extends by a unit length, the surface energy of the glass increases due to the creation of new free surface area, and this increase is balanced by the strain energy release rate $G_c$ released by the glass during the unit-length crack extension. When $G_c$ reaches two times of the surface energy $\gamma$ of the glass, $G_c = 2\gamma$, the crack can propagate. Griffith used Inglis's equation for stress concentration in an infinitely narrow elliptical cavity to derive a formula for the strength of materials [4], as follow

$$\sqrt{2\gamma E'} = \sqrt{G_c E'} = \sigma\sqrt{\pi a} \qquad (1)$$

where $\sigma$ is the stress. 2a is the length of crack.

Irwin [5] and Orowan [6, 7] expounded upon Griffith's work and defined the stress-intensity factor $K$,

$$K = \sqrt{G_c E'} \qquad (2)$$

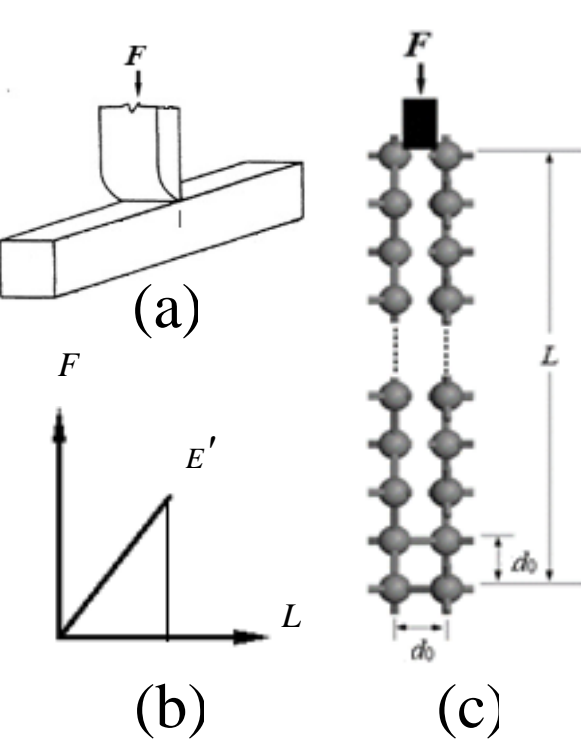

FIG 1. (a) Crack in specimen for the cleavage toughness measurements on the crystal plane. (b) Force–displacement relationship. (c)Loaded condition of micro crack with length $L$, $d_0$ is spacing of planes.

Irwin suggested the critical stress-intensity factor $K$ should be comparable with the fracture toughness $K_C$. The fracture toughness is generally derived from experiments [8]. Using Equation (1) Griffith obtain a calculated value of surface energy of glass of $\gamma$ = 1.75 J/m$^2$. The experimental value of surface energy measured by himself is only 0.54 J/m$^2$ [2, 4]. Such significant error between the theoretical and experimental surface energy demonstrate that Equation (1), which is based on the continuum concept, cannot yield precise results on the atomic scale. Similar problems have also been encountered in other materials. Therefore, scientists are increasingly devoted to study the crack propagation at the atomic level [9-18]. However, up to now, predicting strain energy release rate and surface energy remains elusive

*Contact author: fmgao@tust.edu.cn

[16-18]. In this work, combining Griffith thermodynamic energy balance with cohesive strength model we find that the bond strength and the bond density can characterize the surface energy, the strain energy release rate and the cleavage toughness. As will be seen, our reformulated equations, no adjustable parameters, work very well in practice. Its simplicity helps to open a new insight into the nature of fracture toughness.

The cleavage fracture refers to the breaking of the bond between the atoms of a material when subjected to external force *F*, leading to the propagation of cracks along the internal cleavage plane. From the point of view of crystal bonding, the crack propagation can be regards as the sequential bond-rupture process, as shown Fig. 1. Fig. 1(a) is the crack in specimen for the cleavage toughness measurements on the crystal plane. Fig. 1(b) is the force – displacement relationship [1]. Fig. 1(c) is the loaded condition of micro crack along the cleavage plane. A force *F* acts on the crystal by means of a punch. In present square pseudolattice model of micro crack [9], we assume that there is a bond within the volume $V_b$ ( $V_b = d_0 \times d_0 \times d_0$). The crack region with a length of L is filled with such bond of the bond volume $V_b$. When the sample is forced, the potential energy is increased. For elastic crystals in the absence of any plastic flow, such as cleavage case, the elastic energy per unit volume $U_e$ can be $\frac{\sigma^2}{2E'}$ [1]. To get the total strain energy released, we need to multiply $U_e$ by the volume in which this energy is released. For the cleavage crack in Fig. 1(c) , the volume of the crack region $V = Ld_0t = Nd_0{}^2t$ , where *t* is the thickness, $L = Nd_0$. Therefore, a total strain energy released per-unit thickness is $U = Nd_0{}^2\sigma^2/(2E')$ . When a crack propagates this decrease in strain energy is balanced by the sequential bond-rupture energy. The total surface energy is $2\gamma Nd_0t$, with $\gamma$ the surface energy per unit area. The surface energy per unit thickness is $U_s = 2\gamma Nd_0$. For the potential energy change of the cleavage crack, we can write,

$$\Delta U = U_s - U = 2\gamma Nd_0 - Nd_0{}^2\sigma^2/(2E') \quad (3)$$

According to Griffith energy-balance concept, the first derivative of the potential energy $\Delta U$ is equal to zero. We can obtain

$$\sqrt{2\gamma E'} = \sigma\sqrt{d_0} \quad (4)$$

where σ is the stress. On the other hand, according to the cohesive strength model [1, 2], the surface energy per unit area, $\gamma$, is expressed as

$$\gamma = \frac{E'd_0}{\pi^2} \quad (5)$$

Thus,

$$\sqrt{2\gamma E'} = \frac{\sqrt{2}E'}{\pi}\sqrt{d_0} \quad (6)$$

It should be noted that $E'$ identifies as Young's modulus *E* in simplest case of plane stress. Schultz et al. claimed that the measured fracture toughness of diamond, 3-5 MPa m$^{1/2}$ are one of the highest reliable single crystal fracture toughness values [3]. The cell parameter of diamond is $3.567 \times 10^{-10}$ m. There are 16 C-C bonds in its cell. Its bond volume $V_b$= $3.567^3 \times 10^{-30}/16$ = $2.84 \times 10^{-30}$ m$^3$, and $d_0$ = $1.42 \times 10^{-10}$ m. If taking $E'$ as Young's modulus *E*, the calculated Irwin's critical stress-intensity factor, *K* = 5.9 MPa m$^{1/2}$, which do not fall within the range of experimental fracture toughness values. If taking $E'$as Young's modulus *E*, the calculated surface energy of Si and Ge are 3.54 J/m$^2$ and 3.00 J/m$^2$, respectively. The experimental surface energy of Si and Ge are only 1.23 J/m$^2$ and 1.06 J/m$^2$, respectively [19]. The obvious overestimation of these calculated values demonstrate that the applied stress may be a more complex state stress. $E'$ needs to take a more appropriate value. In complex state of stress, $E'$ may be identified with a function of $f(E, \nu)$. Since elastic materials can be represented by either pair of constants $(E, \nu)$ or $(G, B)$, $E'$can also be expressed as $E' = f(G, B)$, where *B* is bulk modulus and *G* is shear modulus. The bulk modulus and shear modulus play the important role in the fracture strength and the resistance to deformation for all materials, respectively [20]. In order to determine $E'$ we suggest a functional relationship as follow,

$$E' = f(G, B) = \eta G - i(1 - \eta)B \quad (7)$$

where $\eta$ is a mixing factor. Thus

$$E' = \sqrt{\eta^2 G^2 + (1 - \eta)^2 B^2} \quad (8)$$

Equation (8) represents the competition between the fracture strength and the resistance to deformation. At the balance point, $\frac{\mathrm{d}E'}{\mathrm{d}\eta} = 0$, we obtain

$$\eta = \frac{B^2}{G^2 + B^2} \quad (9)$$

$$E' = \frac{GB}{\sqrt{G^2 + B^2}} \quad (10)$$

Using the homogeneous approximation, $B = \frac{E}{3(1-2\nu)}$ and $G = \frac{E}{2(1+\nu)}$, Equation (10) may be written in terms of Young's modulus and Poisson's ratio, as follows

$$E' = \frac{E}{\sqrt{13-28\nu+40\nu^2}} \quad (11)$$

Here, we rewrite Equation (6), and define the cleavage toughness $K_C^G$, as follows

$$\sqrt{\pi^2 \gamma E'} = E'\sqrt{{V_b}^{1/3}} = K_C^G \quad (12)$$

which is different from Irwin's stress-intensity factor in Equation (2). In Equation (2) $G_c$ is the strain energy release rate corresponding to the $K$. By comparing with Equation (2), we may define a pseudo strain energy release rate $G_C^G$, as follows

$$G_C^G = \pi^2 \gamma = E' {V_b}^{1/3} \quad (13)$$

The parameter $G_C^G$ is the strain energy release rate corresponding to the $K_C^G$, and $G_C^G = \frac{\pi^2}{2} G_c$. The bond density $N_A$ is defined as the number of bonds per unit area, $N_A = 1/V_b^{2/3}$. Thus

$$K_C^G = \frac{GB\sqrt{{V_b}^{1/3}}}{\sqrt{B^2+G^2}} = \frac{GB}{\sqrt{(B^2+G^2)N_A^{1/2}}} \quad (14)$$

$$K_C^G = \frac{E}{\sqrt{(13-28\nu+40\nu^2)N_A^{1/2}}} \quad (15)$$

where the unit of the bond density $N_A$ is in $m^{-2}$, the unit of $E$, $G$ and $B$ are in MPa, and the unit of $K_C^G$ is in MPa $m^{1/2}$. The shear and bulk modulus of diamond is 460 GPa and 500 GPa, respectively [21]. Using Equation (14), the calculated cleavage toughness of the covalent crystal diamond is 4.03 MPa $m^{1/2}$, which falls well within the range of experimental toughness values. Our calculated value for ionic crystal NaCl, 0.18 MPa $m^{1/2}$, is also very consistent with the experimental toughness value, 0.17 MPa $m^{1/2}$ [3]. These results demonstrate that $K_C^G$ can characterize the measured fracture toughness. Schultz et al. selected the ionic crystal LiF, the primarily covalent crystal GaP and Si, and the mixed character bond crystal $MgA1_2O_4$ as prototypes to determine the criterion of the cleavage for ionic and covalent crystals, because it is generally accepted that the {100} planes of LiF, the {110} planes of GaP, the {111} planes of Si, and the {100} planes of $MgA1_2O_4$, are the cleavage planes, which possess the lowest measured toughness $K_C$ [3]. Here, we calculated the cleavage toughness of these crystals which are in very good agreement with measured fracture toughness values, as listed in Table I. Therefore, we can conclude that the cleavage toughness is determined by the elastic modulus and the bond density. The constant $K_C^G$ is indeed different with Irwin's stress-intensity factor. Some detractors have argued the rationality of the equivalence relationship between the stress-intensity factor and the fracture toughness [22]. We also noticed that Niu et al. proposed the empirical formula [15]. However, their calculated results of the toughness are significantly larger than experimental ones, as shown in Table I.

The strain energy release rate and the surface energy of cleavage plane also can be calculated by the elastic modulus and the bond density, as follows

$$G_C^G = \frac{GB}{\sqrt{(B^2+G^2)N_A}} \quad (16)$$

$$G_C^G = \frac{E}{\sqrt{(13-28\nu+40\nu^2)N_A}} \quad (17)$$

$$\gamma = \frac{GB}{\pi^2\sqrt{(B^2+G^2)N_A}} \quad (18)$$

$$\gamma = \frac{E}{\pi^2\sqrt{(13-28\nu+40\nu^2)N_A}} \quad (19)$$

TABLE I. Experimental and calculated cleavage toughness for several single crystals. $K_C$ (MPa $m^{1/2}$) is experiment toughness [3]. $K_C^G$ (MPa $m^{1/2}$) is from this work, $K_{Niu}$ (MPa $m^{1/2}$) is from Niu's calculations [15]. $G$ (GPa) is shear modulus, $B$ (GPa) is bulk modulus [21], $V_b$ ($10^{-30}$ $m^3$) is bond volume.

| | LiF | GaP | Si | $MgAl_2O_4$ |
|---|---|---|---|---|
| $V_b$ | 2.73 | 10.12 | 10.01 | 4.12 |
| $B$ | 45 | 56.5 | 66.2 | 109 |
| $G$ | 65 | 89.3 | 97.7 | 197 |
| $K_C$ | 0.50 | 0.65 | 0.82 | 1.18 |
| $K_C^G$ | 0.44 | 0.70 | 0.80 | 1.21 |
| $K_{Niu}$ | 0.77 | 1.2 | 1.3 | 1.9 |

The calculated surface energy $\gamma$ of Si and Ge is 1.20 $J/m^2$ and 1.01 $J/m^2$, respectively. The calculated $\gamma$ of Si and Ge is in good agreement with the experimental values [19]. The pseudo strain energy release rate $G_C^G$ of Si is 12 $J/m^2$, which is significantly higher than Griffith's strain energy release rate $G_c$ , 2.4 $J/m^2$. The value of $G_C^G$ is more comparable to the strain energy release rate when the crack velocity is a limiting crack speed (2/3 of the Rayleigh wave speed) [17, 18].

In summary, the cleavage toughness of single crystals is studied in the framework of Griffith fracture theory. The appropriate elastic modulus of single crystals is found using the complex variable function method, which shows a higher accuracy in practice of toughness calculations than the traditional approximate modulus in case of plane strain, $E/(1-\nu^2)$. The cleavage toughness is expressed as the appropriate elastic modulus by multiply the sixth root of the bond volume, or divided by the fourth root of the bond density. It is applied to predict the toughness of the crystal cleavage plane for typical ionic and covalent single crystals the results are in good agreement with the experimental values. The presented approach of surface energy also works very well in practice. Owing to its simplicity, our approach may be easily extendable to the fracture study of various materials.

The author acknowledges the financial support from the National Natural Science Foundation of China (Grant No. 22379112, 21071122)

---

[1] M. A. Meyers and K.K. Chawla, Mechanical behavior of materials (Cambridge University Press, New York, 2008), 2nd ed.

[2] B. Lawn, Fracture of brittle solids (University Press, Cambridge, UK, 1993), 2nd ed.

[3] R. A. Schultz, M. C. Jensen, and R. C. Bradt, Single crystal cleavage of brittle materials. Int. J. Fract. 65, 291 (1994).

[4] A. A. Griffith, The phenomena of rupture and flow in solids, Phil. Trans. Roy. Soc.Lond. A, 221, 163 (1921).

[5] G. R. Irwin, Trans. ASME, Ser. E, Analysis of stresses and strains near the end of a crack traversing a plate, J. Appl. Mech. 24, 361 (1957).

[6] E. Orowan, Fracture and strength of solids, Rep. Prog. Phys., 12, 185 (1949).

[7] E. Orowan, Energy criteria of fracture, Weld. Res. Supp., 34, 157 (1955).

[8] G.R. Irwin, Fracture testing of high-strength sheet materials under condition appropriate for stress analysis, NRL Rep. No., 5486 (1960).

[9] C. Hsieh, R. M. Thomson, Lattice theory of fracture and crack creep, J. Appl. Phys. 44, 2051 (1973).

[10] J. J. Gilman, Direct measurements of the surface energies of crystals, J. Appl. Phys. 31 2208 (1960).

[11] G. I. Barenblatt, The mathematical theory of equilibrium cracks in brittle fracture, Adv. Appl. Meth., 7, 55 (1962).

[12] H. A. Elliot, An analysis of the conditions for rupture due to Griffith cracks, Proc. Phys. Soc. Lond., 59, 208 (1947).

[13] S. W. King and G. A. Antonelli, Simple bond energy approach for non-destructive measurements of the fracture toughness of brittle materials. Thin Solid Films, 515, 7232 (2007).

[14] L. Borgsmiller, M. T. Agne, J. P. Male, S. Anand, G. Li, S. I. Morozov, and G. J. Snyder, Estimating the lower-limit of fracture toughness from ideal-strength calculations. Mater. Horiz. 9, 825 (2022).

[15] H. Niu, S. Niu, and A. R. Oganov, Simple and accurate model of fracture toughness of solids, J. Appl. Phys. 125, 065105 (2019).

[16] T. Nguyen and D. Bonamy, Role of crystal lattice structure in predicting fracture toughness, Phys. Rev. Lett. 123, 205503 (2019).

[17] J.E. Hauch D. Holland, M.P. Marder, and H. L. Swinney, Dynamic fracture in single crystal silicon, Phys. Rev. Lett. 82, 3823 (1999).

[18] T. Cramer, A. Wanner, and P. Gumbsch, Energy dissipation and path instabilities in dynamic fracture of silicon single crystals, Phys. Rev. Lett. 85, 788 (2000).

[19] R. J. Jaccodine, Surface energy of germanium and silicon, J. Electrochem. Soc. 110, 524 (1963).

[20] S. F. Pugh, Relations between the elastic moduli and the plastic properties of polycrystalline pure metals, Philos. Mag. 45, 823 (1954).

[21] W. J. Tropf, T. J. Harris, and M. E. Thomas, In Electro-optics handbook, edited by R.W. Waynant and M. N. Ediger (McGraw-Hill Press, Columbus, 2000) 2nd ed.

[22] H. C. Liu, Solid fracture research (Defense Industry Press, Beijing, 2013).